\documentstyle [twocolumn,epsf,amssymb,amsmath]{mn}
\newcommand{\mincir}{\raise
-3.truept\hbox{\rlap{\hbox{$\sim$}}\raise4.truept\hbox{$<$}\ }}
\newcommand{\magcisr}{\raise
-3.truept\hbox{\rlap{\hbox{$\sim$}}\raise4.truept\hbox{$>$}\ }}
\newcommand{\minmag}{\raise
-3.truept\hbox{\rlap{\hbox{$<$}}\raise5.truept\hbox{$<$}\ }}
\newcommand{\be}{\begin{equation}}
\newcommand{\ee}{\end{equation}}

\newcommand{\ba}{\begin{eqnarray}}
\newcommand{\ea}{\end{eqnarray}}
\newcommand{\brr}{\begin{array}}
\newcommand{\err}{\end{array}}
\newcommand{\bc}{\begin{center}}
\newcommand{\ec}{\end{center}}

\renewcommand{\baselinestretch}{1.0}

\title[Comparison of the linear bias models in the light of the Dark Energy Survey]
{Comparison of the linear bias models in the light of the Dark Energy Survey}
\author[Alexandros Papageorgiou, Spyros Basilakos, Manolis Plionis] {A. Papageorgiou$^{1,2}$, S. Basilakos$^{2}$, M. Plionis$^{1,3}$\\
\vspace{0.1cm} $^{1}$ Physics Department, University of Thessaloniki, Thessaloniki 54124,Greece \\
  $^{2}$ Academy of Athens, Research Center for Astronomy \& Applied
  Mathematics, Soranou Efessiou 4, 11-527, Athens, Greece \\
$^{3}$ National Observatory of Athens, Lofos Nymfon, 11851 Athens, Greece
  }

\begin{document}
  
\maketitle

\begin{abstract}
The evolution of the linear and scale independent bias, based on
the most popular dark matter bias models within the $\Lambda$CDM
cosmology, is confronted to that of the Dark Energy Survey (DES)
Luminous Red Galaxies (LRGs).
Applying a $\chi^2$ minimization procedure between models and data 
we find that all the considered linear bias models reproduce well the
LRG bias data.
The differences among the bias models are absorbed in the predicted
 mass of the  dark-matter halo in which LRGs live and 
which ranges between 
$\sim 6 \times 10^{12} h^{-1} M_{\odot}$ and $1.4 \times
10^{13} h^{-1} M_{\odot}$, for the different bias models. 
Similar results, reaching however a maximum value of $\sim 2\times
10^{13} h^{-1} M_{\odot}$, are found by confronting the 
SDSS (2SLAQ) Large Red Galaxies clustering with
theoretical clustering models, which also include the evolution of
bias. This later analysis also provides a value of
$\Omega_{m}=0.30\pm 0.01$, which is in excellent agreement with recent
joint analyses of different cosmological probes
and the reanalysis of the Planck data.



\end{abstract}

\vspace{1.0cm}

\section{Introduction}
Studies of the distribution of matter on large-scales, based on 
different mass tracers (galaxies, clusters etc), can be used 
to test the validity of different models of structure formation. 
However, an important issue that significantly affects such an 
approach is our limited knowledge of how 
luminous matter traces the background mass density field. 
The so-called
biasing between different extragalactic sources and the 
underlying matter distribution was first introduced  
(e.g. Kaiser 1984, Bardeen et. al. 1986) in order to explain
the lower amplitude of the 2-point correlation function of galaxies
with respect to that of galaxy clusters.

The most common biasing models consider the Large Scale Structures (LSS)
as high peaks of an initially random Gaussian 
density field, while assuming scale-independence (mostly 
above $5h^{-1}$ Mpc) and linearity.
Following the above lines the linear bias parameter is defined as the
ratio of the fluctuations of the mass 
tracer ($\delta_{tr}$) to that of the underlying mass ($\delta_{m}$):

\begin{equation}\label{eq:wp1}
 b = \frac{\delta_{tr}}{\delta_{m}}\;.
\end{equation}

Due to the fact that the two point correlation function is written as 
$\xi(r)=\langle\delta(x)\delta(x+r)\rangle$, one can easily show that the 
bias factor is also given by: 
\begin{equation}\label{eq:wp2}
b = \left(\frac{\xi_{tr}}{\xi_{m}}\right)^{1/2}
\end{equation}
or 
\begin{equation}\label{eq:wp3}
b = \left(\frac{\sigma_{8,tr}}{\sigma_{8,m}}\right)^{1/2}
\end{equation}
where $\sigma^{2}_{8,i}=\xi_{i}(0)=\langle\delta^2_{i}(x)\rangle$ is 
the 
mass variance at $R_{8}=8 h^{-1}$ Mpc ($i$ corresponding to $tr$ or
$m$).

In the literature there is a large body of scenarios that 
attempt to predict the cosmological evolution of the bias parameter
and in general, there are two main categories of analytic bias 
evolution models (for more details see Papageorgiou et. al. 2012 
and the references therein). 
The first family of models corresponds to the so-called 
\emph{galaxy merging} bias which  
is based on the Press-Schechter formalism (1974), the peak 
background split \cite{Bard1986} and the spherical 
collapse model (Cole \& Kaiser 1989; Mo \& White 1996; 
Matarrese et. al. 1997; Moscardini et. al. 1998; Sheth \& Tormen 1999;
Valageas 2009, 2011).
The difference between the predictions of these bias models 
with respect to those of numerical simulations 
have led the authors to introduce modifications to the models by using 
ellipsoidal collapse \cite{SMT2001}, 
new fitting bias formulas (Jing 1998; Tinker et. al. 2005) and 
a non-Markovian extension of the excursion set 
theory (Ma et. al. 2011; de Simone et. al. 2011).

The second family assumes a continuous mass-tracer fluctuation field,
which is proportional to that of the underlying 
mass. In this framework, the mass tracers act as ``test particles''. 
This family can be divided into two sub-groups:

\begin{itemize}
\item The first one is the \emph{galaxy conserving} bias model. 
This model utilizes the continuity equation 
and the assumption that the extragalactic tracers and underlying mass 
share the same velocity field (Nusser \& Davis 1994; Fry 1996; Tegmark
\& Peebles 1998; Hui \& Parfrey 2008; Schaefer, Douspis \& Aghanim
2009). In this context, the evolution of bias  
is given by $b(z)=1+(b_{0}-1)/D(z)$, 
as the solution of a 1st order differential equation. Notice, that
$D(z)$ is the growth factor of density perturbations 
(scaled to unity at the present time)
and $b_{0}$ is the bias factor at the present epoch.
It is well known that this bias model 
has two fundamental problems. The first one is related with the 
anti-biased problem, the fact
that an anti-biased set of tracers at the present time ($b_{0}<1$) 
remains always anti-biased at high redshifts. The second problem is
based on the fact that this model shows a realistic bias evolution
only at low redshifts $z \le 0.5$ \cite{Bagla1998}. 

\item This subfamily is basically an extension of 
the previous one but free of the above problematic issues.
Specifically, it utilizes the three hydrodynamical equations of motion 
(continuity, Euler and Poisson equations) in the linear regime 
and the fact that the underlying mass and extragalactic mass-tracers 
feel the same gravity field, but they do not not necessarily share the same 
velocity field. The combination of the above ingredients provide 
a second order differential equation in
$b$, the solution of which gives the evolution of the
linear bias factor (Basilakos \& Plionis 2001, 2003;  Basilakos,
Plionis \& Ragone-Figueroa 2008; Basilakos, Plionis \& Pouri 2011). It is
interesting to mention that this bias formula 
is valid for all dark energy models including those
of modified gravity (see Basilakos, Plionis \& Pouri 2011;
Basilakos et. al. 2012).
\end{itemize}
It should be mentioned that
the linear bias model relates 
a mass tracer, 
being a galaxy, an Active Galactic Nuclei (AGN), a 
Luminous Red Galaxy (LRG) or a cluster of galaxies, with a host
dark matter halo within which the mass tracer forms and
evolves. The models themselves follow the linear evolution 
of the host halo and not the internal evolution of the
astrophysical processes of the tracer. Thus the assumption 
is that the effects of nonlinear gravity and 
hydrodynamics (merging, feedback mechanisms, etc.) can be
ignored in the linear-regime (Catelan et al. 1998) and that each
DM halo hosts only one mass tracer.

Recently, the deterministic and linear nature of bias has been
challenged (see McDonald et. al. 2009; Chan et al. 2012), and indeed 
a large body of papers 
have been published studying the evolution of bias in the non-linear and 
non-local regimes respectively
(Paranjape et. al. 2013; Assasi et. al. 2014; Di Porto et. al. 2016; 
Lazeyras et. al. 2016; Desjacques et. al. 2016 and references therein).
Morever, numerical simulations have been used 
in several studies 
(Hoffmann et. al. 2017, Hoffmann et. al. 2015, Baldauf et. al. 
2012; Bel et. al. 2015)
towards investigating 
the nature of bias and they found deviations from the linear regime.
Despite  the  above  considerations,  the  linear  biasing
assumption has a long history in cosmology and it 
is still a useful first-order approximation which,
because of its simplicity, is used
in studies of large-scale (linear) dynamics. For example, 
the Dark Energy Survey (DES) team (Elvin-Poole et al. 2017)
used the clustering properties of the 
Luminous Red Galaxies (LRGs) in order to measure the evolution of bias 
in the linear regime.


In the present paper we investigate the predictions of the most
popular of the above linear bias models. Utilizing the 
linear bias data of Luminous Red Galaxies (LRGs), 
recently released by the DES group but also of the
SDSS DR5 data, 
we test the range of validity of the explored bias models. 
In particular, the outline of this paper is as follows. 
In Section 2 we provide the LRGs bias data based on the 1-year DES sample. 
In Section 3 we introduce the main elements of the 
bias evolution models. In Section 4 we present the outcome of the 
current analysis, while in Section 5 we 
compare the DES results with those of SDSS DR5 LRGs data, using 
the 2 point angular correlation function data. 
In Section 6 we provide our conclusions. 

Finally, we would like to spell out clearly which are
the basic assumptions of our work, which are common
to many studies of the bias: 
(a) the biasing is linear on the scales of
interest (which does not preclude being scale dependent
on small non-linear scales), and (b) each dark matter
halo is populated by one extragalactic mass tracer (in our case LRG).

\section{DESY1 Red Galaxies Bias Data}
The application of the correlation function analysis on samples of 
high redshift extragalactic sources for cosmological studies has a long history 
(cf. Basilakos 2001; Matsubara 2004). 
For example, in a sequence of previous publications some of us used the 
clustering properties of the XMM-Newton X-ray sources in order to place constraints 
on the dark energy models (Basilakos \& Plionis, 2005, 2006, 2009, 2010).

In the current work we will use as tracers of the LSS 
the Luminous Red Galaxies (LRGs), which 
according to Eisenstein et. al. (2001), and due to their
high luminosities, are very useful tracers of the LSS. One
important advantage is that such a population can be observed up to
relatively large redshifts. In particular, 
we will use the 
DES bias data provided by Elvin-Poole et al. (2017).
These bias data were extracted with the aid of 
the angular correlation function (ACF), which was
estimated in Elvin-Poole et. al. (2017),  
using the 1-year DES sample 
of $\sim 6.6\times10^{5}$ LRGs in the redshift range $0.15<z<0.9$.

These authors used the assumption of linear bias 
in the derivation of the DES bias data. 
According to Krause et. al. (2017) the scale of $\sim 8h^{-1}$Mpc, 
used by the DES team, ensures that 
the impact of non-linear effects on clustering and thus on biasing 
is almost negligible.
Moreover, in order to measure the DES bias the above authors 
have fixed the cosmological parameters at the mean of 
the so called DESY1COSMO 
posterior, namely $\Omega_{m}=1-\Omega_{\Lambda}=0.276$, $h=0.7619$, $\Omega_{b}=0.0526$ and 
spectral index $n=0.9964$.\footnote{
We would like to point that  
the DES paper of 
Elvin-Poole et al. (2017) have a typo (Elvin-Poole private 
communication). Specifically, in the first draft of 
{\it arXiv:1708.01536}  
one may see that $\Omega_{m}=0.2276$,
but according to Elvin-Poole the correct value is $\Omega_{m}=0.276$.
Based on the latter value 
the mass variance at $8h^{-1}$Mpc, 
is $\sigma_{8}\simeq 0.8296$.
Also, for the comoving distance and for the dark matter halo mass we
use the traditional parametrization $H_{0}=100h$km/s/Mpc, hence the units 
of the above are given in $h^{-1}$Mpc and $h^{-1}M_{\odot}$. 
Of course, when we compute the power spectrum shape 
parameter $\Gamma$ we use the exact value of $h$.}

Although we will utilize the bias data provided by the previous
references, for completion we rescale the bias data using 
different references cosmologies.  
Specifically, we wish to convert the value of bias data 
from the DESY1COSMO cosmological model, say DES, to
another, say A. Utilizing the definition of bias and the notations of 
Papageorgiou et. al. (2012) the scaling relation from the DESY1COSMO model 
to that of A, takes the form:
\begin{equation}\label{eq:w5}
b_A(z)\simeq b_{B}(z) \frac{\sigma_{8,B}}{\sigma_{8,A}}\frac{D_B(z)}{D_A(z)} \\
\end{equation}
where the index $B$ corresponds to DES, 
$\sigma_{8}$ is the present value of the mass variance at $8h^{-1}Mpc$
and $D(z)$ is the growth factor of matter fluctuations in the linear regime. 
Also, the normalized Hubble parameter of the $\Lambda$CDM model
is given by 
\begin{equation}
E(z)=\left[\Omega_{m}(1+z)^{3}+\Omega_{\Lambda}\right]^{1/2} \;.
\end{equation}
where $\Omega_{\Lambda}=1-\Omega_{m}$.

In the current article we convert the bias data to the following 
basic $\Lambda$CDM cosmologies:
\begin{itemize}
  \item Planck TT+lowP+lensing results, namely 
$\Omega_{m}=1-\Omega_{\Lambda}=0.308$, $h=0.678$, $n=0.9671$, $\Omega_{b}=0.0484$ 
and $\sigma_{8}=0.815$. The latter cosmological parameters 
are in agreement with those of the reanalysis of the Planck
data provided by Spergel et al. (2015). 

\item Finally, we utilize the DES/Planck/JLA/BAO
joint likelihood results, namely  
$\Omega_{m}=1-\Omega_{\Lambda}=0.301$, $h=0.682$, $\Omega_{b}=0.048$, $n=0.973$ 
and $\sigma_{8}=0.801$ (see Abbott et al. 2017). 
\end{itemize}

In Table \ref{tab01} we present the precise numerical
values of the DES bias data with the corresponding errors that are
used in our analysis. Also, in the last two columns of Table 1 
we provide with the aid of Eq.(\ref{eq:w5})
the Planck-scaled and DES/Planck/JLA/BAO-scaled bias data respectively.

\begin {table*}
\begin{center}
    \begin{tabular}{ c c  c  c c}
    \hline 
    \\[0.1ex]
Red. Range&    Median Redshift     & DESY1 bias         & Planck-scaled bias    & DES/Planck/JLA/BAO-scaled bias                               \\[1.ex]  \hline 
$0.15<z<0.3$   & $0.225\pm 0.075$ &  $1.40\pm0.077$   &  $1.434\pm0.079$  &        $1.457\pm0.080$ \\     
$0.3<z<0.45$   & $0.375\pm 0.075$ &  $1.61\pm0.051$   &  $1.655\pm0.052$  &        $1.680\pm0.053$     \\
$0.45<z<0.6$   & $0.525\pm 0.075$ &  $1.60\pm0.040$   &  $1.649\pm0.041$  &        $1.673\pm0.042$    \\ 
$0.6<z<0.75$   & $0.675\pm 0.075$ &  $1.93\pm0.045$   &  $1.994\pm0.047$  &        $2.022\pm0.047$     \\ 
$0.75<z<0.9$   & $0.825\pm 0.075$ &  $1.99\pm0.066$   &  $2.060\pm0.068$  &        $2.088\pm0.069$     \\  \hline 
    \end{tabular}
\end{center}
\caption {The measured bias data 
of the 1-year DES LRGs from Elvin-Poole et. al. (2017).}
\label{tab01}
\end{table*}

\section{Bias Models}
In this section we briefly present the most popular bias models. 
Specifically, from the galaxy merging bias family we will investigate the models of Sheth, Mo \& Tormen (2001)
[hereafter SMT], the Jing (1998) the Tinker et. al. (2010) [hereafter TRK], de Simone et. al. (2011) [hereafter DMR] 
and the Ma  et. al. (2011) [hereafter MMRZ]. 
In this case the bias
factor is given as a function of the peak-height parameter, $\nu =
\delta_{c}(z)/\sigma(M_h,z)$ where $\delta_{c}$ is the linearly
extrapolated density threshold above which structures collapse. 
In the present study we utilize the accurate fitting formula 
of Weinberg \& Kamionkowski (2003) to estimate $\delta_{c}(z)$.
Furthermore, the mass variance is written as 

\begin{equation}\label{eq:wp18}
\sigma(M_h,z)=\left[\frac{D^2(z)}{2\pi^2}\int_0^{\infty}k^2P(k)W^2(kR)dk\right]^{1/2}
\end{equation}
where $W(kR) = 3[sin(kR) - kRcos(kR)]/(kR)^3$ is 
the Fourier transform of the top-hat smoothing kernel with 
$R = [3M_h/(4\pi\rho_{m})]^{1/3}$, $M_h$ is the mass of the halo 
and $\rho_{m}$ is the mean matter density of the universe at the present time.
The quantity $P(k, z)$ is the CDM linear power spectrum given by $P(k, z)=P_0k^nT^2(k)D^2(z)$ where 
$n$ is the spectral index of the primordial power 
spectrum and $T(k)$ is the CDM transfer function 
provided by Eisenstein \& Hu (1998):

\begin{equation}\label{eq:wp16}
T(k) = \frac{L_0}{L_0 +C_0q^2}
\end{equation}
with $L_0={\rm ln}(2e+1.8q)$, $e=2.718$, $C_0 = 14.2+\frac{731}{1+62.5q}$ 
and $q =k/\Gamma$ with $\Gamma$ being 
is the shape parameter 
given by (Sugiyama 1995):
$$
\Gamma= \Omega_{m}h{\rm exp}(-\Omega_{b}-\sqrt{2h}\frac{\Omega_{b}}{\Omega_{m}}).
$$
Taking the aforementioned quantities into account and using 
Eq.(\ref{eq:wp18}) the normalization of the power spectrum becomes
\begin{equation}\label{eq:wp19}
P_0 = 2\pi^2\sigma_8^2\left[\int_0^{\infty}T^2(k)k^{n+2}W^2(kR_8)dk\right]^{-1}
\end{equation}
where $\sigma_{8}\equiv \sigma(R_{8},0)$.

From the second bias group we will use the 
generalized model of Basilakos, Plionis \& Pouri 2011 (hereafter BPR;
see also Basilakos et. al. 2012) which is 
valid for any dark energy model 
including those of modified gravity.

Let us now briefly present the functional forms of the aforementioned 
linear bias models (for more details see Papageorgiou et al. 2012 and
references therein), whose  dark matter halo masses can be constrained
by using the DES bias data:

\begin{description}
  \item[\bf SMT:] 
  \begin{equation}\label{eq:wp26}
b(\nu) = 1 + \frac{1}{\sqrt{\alpha}}\delta_c(z)[\sqrt{\alpha}(\alpha\nu^2)
+ \sqrt{\alpha}b(\alpha\nu^2)^{1-c} - f(\nu)]
\end{equation}
with
\begin{equation}\label{eq:wp27}
f(\nu) = \frac{(\alpha\nu^2)^c}{(\alpha\nu^2)^c + b(1-c)(1 - c/2)},
\end{equation}
where $\alpha = 0.707, b = 0.5, c = 0.6$.

  \item[\bf JING:] 
  \begin{equation}\label{eq:wp101}
b(\nu) = \left(\frac{0.5}{\nu^4}+1\right)^{0.06-0.02\nu}\left(1+\frac{\nu^2-1}{\delta_c}\right)
\end{equation}

  \item[\bf TRK:]
  \begin{equation}\label{eq:wp28}
b(\nu) = 1 - A \frac{\nu^{\alpha}}{\nu^{\alpha} + \delta_c^{\alpha}} + B \nu^b + 
C\nu^c \;,
\end{equation}
where $A = 1+0.24y$exp$[-(4/y)^4]$, $B = 0.183$, $C = 0.019 + 0.107y + 0.19$exp$[-(4/y)^4]$, $\alpha = 0.44y - 0.88$, $b=1.5$ and $c = 2.4$. For $y$ we have $y = {\rm log}_{10}\Delta$.

  \item[\bf DMR:]
  
  \begin{multline}\label{eq:wp200}
b(\nu) = 1 + \sqrt{\alpha}\frac{\nu^2}{\delta_c}\left[1+0.4\left(\frac{1}{\alpha\nu^2}\right)^{0.6}\right] \\
-\frac{1}{\sqrt{\alpha}\delta_c\left[1+0.067\left(\frac{1}{\alpha\nu^2}\right)^{0.6}\right] } \;,
\end{multline}
where $\alpha = 0.818$.

   \item[\bf MMRZ:]
   \begin{equation}\label{eq:wp100}
b(\nu) = 1 + \frac{\alpha\nu^2-1+\frac{\alpha\kappa}{2}\left[2-e^{\alpha\nu^2/2}\Gamma(0,\frac{\alpha\nu^2}{2})\right]}{\sqrt{\alpha}\delta_c[1-\alpha\kappa+\frac{\alpha\kappa}{2}e^{\alpha\nu^2/2}\Gamma(0,\frac{\alpha\nu^2}{2})]}\;,
\end{equation}
where $\alpha = 0.818$ and $\kappa = 0.23$.

   \item[\bf BPR:]
   
   \begin{equation}\label{eq:wp23}
 b(z) = 1 + \frac{b_0 - 1}{D(z)} + C_2 \frac{J(z)}{D(z)}
\end{equation}
where
\begin{equation}\label{eq:wp24}
b_0 = 0.857\left[1+\left(\frac{\Omega_{m}}{0.27}\frac{M_h}{10^{14}h^{-1}M_{\odot}}\right)^{0.55}\right]
\end{equation}
and
\begin{equation}\label{eq:wp25}
C_2 = 1.105\left(\frac{\Omega_{m}}{0.27}\frac{M_h}{10^{14}h^{-1}M_{\odot}}\right)^{0.255}
\end{equation}
Notice that the factor $\Omega_{m}/0.27$ comes from the fact that 
the constants $b_0$ and $C_2$ where originally computed (Basilakos et al. 2012)
using $\Lambda$CDM N-body simulations in the context of WMAP7 cosmology, namely 
$\Omega_m =0.27$ and $\sigma_{8}=0.81$. Interestingly, this $\sigma_8$
value is consistent with the most recent Planck analysis of Ade et al. (2015). 

\end{description}

\section{Fitting Models To The Bias Data}
In order to test the range of validity 
of the aforementioned bias models
we use a standard 
$\chi^2$-minimization procedure and compare the 
measured LRG bias data \cite{DES2017} with the 
expected theoretical bias models. In our case 
the $\chi^2$ function is defined as follows:

\begin{equation}
\chi^2 = \displaystyle\sum^{5}_{i=1}\frac{\left[b_{\rm obs}(z_{i})-
b_{\rm th}(\textbf{p},z_{i}) \right]^2}{\sigma_{i}^{2}+\sigma_{z}^{2}}
\end{equation}
where ${\bf p}$ is a vector containing the 
free parameter that
we want to constrain. Also, 
$\sigma_{i}$ is the uncertainties of the observed bias 
(see Table 1). 
The fact that the DES bias data are given in redshift 
intervals implies that 
we need to introduce an additional uncertainty in the $\chi^{2}$ estimator.
In our case we choose this uncertainty to be equal to the width
of the redshift bin, $\sigma_{z}=0.075$.
It becomes clear that our statistical analysis
contains one independent free parameter, hence
the statistical vector 
${\bf p}$ is associated with the environment of the dark matter halo in which
the mass tracers (in our case LRGs galaxies) live, 
namely ${\bf p}=M_{h}$.

To this end we utilize, the corrected
Akaike information criterion which is appropriate for small sample size,
(Akaike 1974, Sugiura 1978). Considering Gaussian errors the 
AIC$_{c}$ estimator becomes (see Liddle 2007)
\begin{equation}\label{AIC}
{\rm AIC}_{c}=\chi^{2}_{\rm min}+2k+\frac{2k(k+1)}{N-k-1}
\end{equation}
where $N$ is the number of data (5 in our case), $k$
is the number of free parameters, and thus when
$k= 1$ then AIC$_{c}=\chi^{2}_{\rm min}+10/3$.
A smaller value of AIC points a better model-data fit. 
Moreover, in order to explore, the effectiveness of the 
different models in reproducing the observational data, we need to 
introduce the model pair difference, namely
$\Delta$AIC=AIC$_{c,x}$-AIC$_{c,y}$. Therefore, the higher the value of
$|{\rm \Delta AIC}|$, the higher the evidence against 
the model with higher value of AIC$_{c}$, with a difference
$4\le |{\rm \Delta AIC}| \le 7$ 
(Burnham \& Anderson 2002; Burnham \& Anderson 2004) 
suggesting a positive
such evidence and
$|{\rm \Delta AIC}| \ge 10$
suggesting a strong such evidence. Notice, that 
if $|{\rm \Delta AIC}| \le 2$ then this is a indication of 
consistency among the two comparison models

Our main statistical results are presented in Table 2, where we
quote the fitted halo mass with the corresponding $1\sigma$ 
uncertainties 
and the goodness of fit statistics ($\chi^{2}$ and AIC$_{c}$), 
for three different expansion models (see section 2).
After considering the best-$\chi^{2}$ 
and the value of the Akaike information criterion we find that 
most bias models 
fit at a statistically acceptable level the DES 
bias data. 
The best model is the SMT, while 
we find a tension between the MMRZ model and the bias data,  
$|\Delta$AIC$|=|{\rm AIC}_{c,\rm SMT}-{\rm AIC}_{c,\rm MMRZ}|>4$.
We observe that the BPR and the TRK bias models predict 
consistent values (within $1\sigma$) of dark matter halos 
with that of SMT.
Lastly, the fact $|\Delta$AIC$|\le 2$ implies that 
the SMT bias model is statistically equivalent with those 
of JING, TRK and DMR models, regards-less the value of the fitted 
DM halo. It becomes evident that the differences of
the bias models are absorbed in the fitted value of the dark-matter halo
mass in which LRGs inhabit, 
and which ranges 
from $\sim 5.9 \times 10^{12}h^{-1} M_{\odot}$ to $1.41 \times
10^{13} h^{-1} M_{\odot}$, for the different bias 
models and in the case of DESY1COSMO bias data.

In order to provide a robust model average value of the
DM halo mass, that hosts LRGs, we utilize an inverse-AIC$_c$
weighting of the different model results. 
We find that the weighted model average and the combined weighted
standard deviation of the DM halo mass are:

\begin{equation}
{\bar M}_{h} = \frac{\sum{w_iM_{h,i}}}{\sum{w_i}}= 1.02 \times 10^{13}h^{-1}M_{\odot}
\end{equation}
and
\begin{equation}
\sigma_{M_h} = \sqrt{\frac{{\sum{w_i(M_{h,i}-{\bar M}_{h})^2}}}{{\sum{w_i}}}} = 0.27 \times 10^{13} h^{-1}M_{\odot}
\end{equation}

Using Eq.(\ref{eq:w5}) to rescale the bias data to the
Planck $\Lambda$CDM (TT+lowP+lensing) cosmology, we obtain a
 DM halo mass that lies in the range $\sim 0.6-1.0 \times 10^{13}h^{-1}
M_{\odot}$ for DMR, JING and MMRZ and  $\sim 1.14-1.46 \times
10^{13} h^{-1} M_{\odot}$ for SMT, TRK and BPR.
The model inverse-AIC$_c$ weighted halo mass is:
$${\bar M}_{h} =1.06 (\pm 0.28)\times 10^{13} h^{-1}M_{\odot}.$$

For the DES/Planck/JLA/BAO $\Lambda$CDM cosmology we find 
$0.58 \times 10^{13}h^{-1} M_{\odot}< M_{h} < 1.1 \times
10^{13} h^{-1}M_{\odot}$ for DMR, SMT, JING and MMRZ and 
$\sim 1.40 \times 10^{13} h^{-1} M_{\odot}$ for TRK and BPR respectively. 
Also, here we have 
$${\bar M}_{h}=1.03 (\pm 0.30)\times 10^{13}h^{-1}M_{\odot} \;.$$

Overall, we see that JING, TRK, MMRZ and BPR bias models provide 
consistent values (within $1\sigma$) of the mass of DM halos
hosting LRGs with that of SMT. Also, we find that  regardless the
value of the fitted DM halo mass, the bias model of SMT 
is statistically equivalent to those of JING, TRK and DMR models,
since $|\Delta$AIC$|\le 2$.
Lastly, we observe that in all cases the inverse-AIC$_c$ weighted mean
of the DM halo mass very close to that of SMT.

In the context of TRK and BPR bias models within the  
Planck (TT+lowP+lensing) and the
DES/Planck/JLA/BAO cosmology respectively, 
it is interesting to mention that 
rescaled bias data provide an LRG host DM halo mass consistent 
at $\sim 2\sigma$ level with that of Sawangwit et al. (2011), namely
$M_{h}\simeq 2(\pm 0.1) \times 10^{13}h^{-1}M_{\odot}$ 
(see also 
Pouri, Basilakos \& Plionis et. al. 2014) 
for $\Omega_{m}\simeq 0.3$.

\begin{table*}
\caption[]{Statistical results for the bias data (see Table 1): The
  $1^{st}$ column indicates the expansion model (see section 2), the
  $2^{nd}$ column corresponds to bias models appearing in section 3
  and the $3^{rd}$ provides the fitted DM halo mass. The remaining
  columns present the goodness-of-fit statistics $\chi^{2}_{\rm min}$,
  AIC$_{c}$ and $|\Delta$AIC$|=|{\rm AIC}_{c,\rm SMT}-{\rm
    AIC}_{c,y}|$. Notice that the index $y$ corresponds to the
  indicated comparison model.}

\tabcolsep 4.5pt
\vspace{1mm}
\begin{tabular}{cccccc} \hline \hline
$\Lambda$CDM Expansion Model & Bias Model & $10^{13}h^{-1}M_{\odot}$ &
  $\chi_{\rm min}^{2}$ &${\rm AIC_{c}}$&
  $|\Delta$AIC$|$ \vspace{0.05cm}\\ \hline
DESY1COSMO (Elvin-Poole et al. 2017) &SMT& $1.090^{+0.179}_{-0.164}$   &2.663& 5.997 & 0   \vspace{0.01cm}\\
            &JING& $0.872^{+0.124}_{-0.115}$   &3.805  & 7.139   & 1.142   \vspace{0.01cm}\\
            &TRK&  $1.409^{+0.194}_{-0.182}$   &3.605  & 6.938   & 0.941   \vspace{0.01cm}\\
            &MMRZ& $0.992^{+0.118}_{-0.111}$   &7.123  & 10.456  & 4.459   \vspace{0.01cm}\\
            &DMR&  $0.594^{+0.100}_{-0.091}$   &2.751  & 6.084   & 0.087   \vspace{0.01cm}\\
            &BPR&  $1.244^{+0.243}_{-0.218}$   &4.975  & 8.308   & 2.311   \vspace{0.15cm}\\
Planck TT+lowP+lensing (Ade et al. 2016)&SMT& $1.148^{+0.185}_{-0.169}$   &2.846& 6.180 & 0   \vspace{0.01cm}\\
            &JING& $0.897^{+0.126}_{-0.116}$   &4.241& 7.574 & 1.394   \vspace{0.01cm}\\
            &TRK& $1.461^{+0.197}_{-0.185}$   &4.064& 7.397 & 1.217   \vspace{0.01cm}\\
            &MMRZ& $1.005^{+0.119}_{-0.111}$   &7.918& 11.251 & 5.071   \vspace{0.01cm}\\
            &DMR& $0.618^{+0.102}_{-0.093}$   &2.967& 6.300 & 0.120   \vspace{0.01cm}\\
            &BPR& $1.293^{+0.237}_{-0.214}$   &5.075& 8.409 & 2.229   \vspace{0.15cm}\\
DES/Planck/JLA/BAO (Abbott et al. 2017) &SMT& $1.094^{+0.173}_{-0.159}$   &2.927& 6.260 & 0   \vspace{0.01cm}\\
            &JING& $0.847^{+0.117}_{-0.109}$   &4.363& 7.696 & 1.436   \vspace{0.01cm}\\
            &TRK& $1.382^{+0.184}_{-0.173}$   &4.214& 7.547 & 1.287   \vspace{0.01cm}\\
            &MMRZ& $0.942^{+0.111}_{-0.104}$   &8.025& 11.358 & 5.098   \vspace{0.01cm}\\
            &DMR& $0.587^{+0.096}_{-0.087}$   &3.053& 6.387 & 0.127   \vspace{0.01cm}\\
            &BPR& $1.421^{+0.253}_{-0.231}$   &5.418& 8.751 & 2.491   \vspace{0.01cm}\\
\hline\hline
\label{tab:growth1}
\end{tabular}
\end{table*}

In order to visualize the behavior of the current bias models against
the data we plot in Fig. 1 the bias evolution models (different
lines), utilizing 
the best fit parameter values given in Table 2. In Fig. 2 we plot the
bias evolution of the different models but when using the Planck
(TT+lowP+lensing)-scaled LRG bias data, while in Fig. 3 we plot the
corresponding curves for the DES/Planck/JLA/BAO-scaled bias data.


Below, we will compare the above results, based on the Dark Energy
Survey LRGs, with those of the SDSS DR5 in order to provide a
relatively complete study regarding the DM halos in which LRGs are embedded.

\begin{figure}
\mbox{\epsfxsize=8.2cm \epsffile{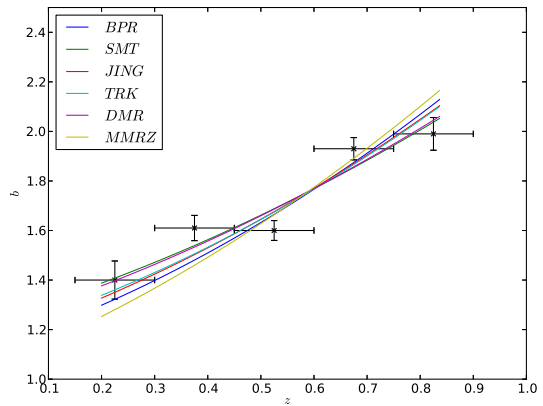}} \caption{Comparison
of the DESY1COSMO bias data with the bias models fits.}
\label{f1}
\end{figure}

\begin{figure}
\mbox{\epsfxsize=8.2cm \epsffile{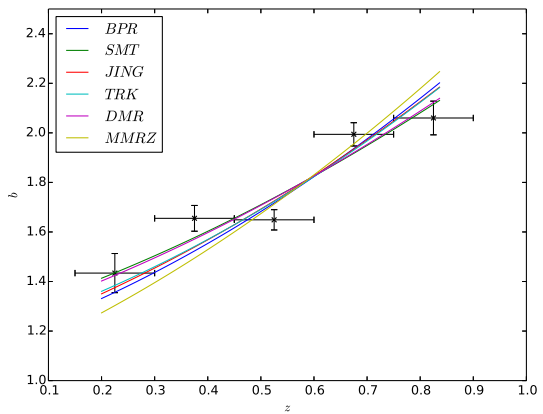}} \caption{Comparison
of the DES bias data scaled to the Planck 
(TT+lowP+lensing) priors with the bias models fits.}
\label{f3}
\end{figure}

\begin{figure}
\mbox{\epsfxsize=8.2cm \epsffile{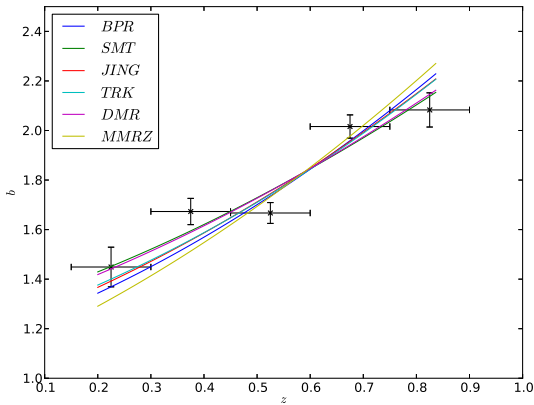}} \caption{Comparison
of the DES  bias data scaled to DES/Planck/JLA/BAO priors with the bias models fits.}
\label{f5}
\end{figure}

\section{Comparison with LRGs from SDSS DR5}
\subsection{Angular Correlation Function Data}
In this section we use the angular correlation function (ACF) of LRGs,
already estimated in Sawangwit et. al. (2011), and compare it with the
theoretical expectations of the $\Lambda$CDM model also incorporating the
effects of the different bias models.

Specifically, we utilize the ACF of $\sim 655775$ photometrically selected LRGs
from the SDSS DR5 catalogue with median redshift $z_{\star} = 0.55$.
This sample has been compiled using the same selection criteria as the
2dF-SDSS LRG and Quasar survey (hereafter 2SLAQ), which covers the
redshift range $0.45<z<0.8$.
Obviously, there is a substantially overlapping with the redshift 
range of the Dark Energy Survey, namely $0.15<z<0.9$.
Based on the original work of Sawangwit et. al. (2011) we utilize the ACF
up to an angular scale of $6000^{''}$ in order to avoid the effects of
BAO's. Since the goal of the present work is to test the performance of 
the most popular linear bias models we also exclude
small angular scales ($\theta<140^{''}$, which translates to $< 1
\; h^{-1}$ Mpc at $z_{\star}$) and for which strong non-linear
effects are expected. However,  
in our theoretical modeling we have taken into account
a non-linear correction as far as the power spectrum is concerned the so called halo-fit model (see below).
Notice, that the precise numerical
values of the ACF data points with the corresponding errors can be found in 
Pouri et al. (2014), while for the total ACF data-set one may check the article 
of Sawangwit et. al. (2011). 


\subsection{Theoretical Angular Correlation Function}
It is well known that the angular correlation function, $w(\theta)$ 
for small $\theta$'s is written as 
(cf. Basilakos \& Plionis 2009 and references therein):

\begin{equation}\label{eq:wp4}
w(\theta)=\frac{1}{2\pi}\int_{0}^{\infty}k^2P(k)dk 
\int_{0}^{\infty}D^2(z)j(k,z,\theta)dz
\end{equation}
with
\begin{equation}\label{eq:wp9}
j(k,z,\theta)=\frac{H_o}{c}\left(\frac{1}{N}\frac{dN}{dz}\right)^2
b^{2}(z)E(z)J_o(k \theta x(z))\;,
\end{equation}
where $x(z)$ is the comoving distance 
\begin{equation}\label{eq:wp12}
x(z) = \frac{c}{H_o}\int_0^z\frac{dy}{E(y)}
\end{equation}
and $J_o$ is the 0th order Bessel function of the first kind given by:
\begin{equation}\label{eq:wp10}
J_0 (\omega)= \frac{1}{\pi}\int^{\pi}_0 \cos(\omega sin\tau)d\tau
\end{equation}
The quantity $1/N$ $dN/dz$ is the normalized source redshift 
distribution, estimated by the fitting formula 
of Pouri et. al. (2014)
\begin{equation}\label{eq:wp5}
\frac{dN}{dz}\propto \left(\frac{z}{z_*}\right)^{\alpha+2}e^{-(z/z_*)^{\beta}}
\end{equation}
with $(z_{\star},\alpha,\beta)=(0.55,-15.53,-8.03)$.
For the power spectrum we are using the nonlinear
power spectrum of Takashi et. al. (2011). Briefly, 
the latter approach 
consists the so called one-halo and two halo terms. 
The first one dominates at small scales, while the 
second one plays a key role at large scales.



\subsection{Fitting models to the LRGs SDSS DR5 ACF data}
In order to quantify the free parameters of the models
we perform a 
$\chi^2$-minimization procedure and compare the measured LRG angular 
correlation function of Sawangwit et al. (2011) with the 
expected theoretical ACF given by 
Eq.(\ref{eq:wp4}). 	
In examining the model dependence we restrict our analysis to 
flat $\Lambda$CDM with 
$n=0.9671$, $\Omega_{b}=0.0484$ 
(Spergel et al. 2015) and vary $\Omega_{m}h$ and $M_{h}$.
Also, in order to treat the $\sigma_{8}-\Omega_{m}$ relation  
we use the following parametrization 
$\sigma_{8}=0.818(0.3/\Omega_{m})^{0.26}$ (Spergel et al. 2015). 
Therefore, the $\chi^2$ function is defined as:

\begin{equation}
\chi^2 = \displaystyle\sum^{11}_{i=1}\left[\frac{w_{obs}(\theta_i)-w_{th}(\theta_i,\textbf{p})}{\sigma_{w,i}}\right]^2
\end{equation}
where $\sigma_{w,i}$ is the uncertainty of the observed ACF 
(see Table 1 in Pouri et al. 2014). Here the statistical vector 
${\bf p}$ contains two independent free 
parameters, namely ${\bf p}=(\Omega_{m}h,M_{h})$. 

\begin{figure*}
\mbox{\epsfxsize=8.2cm \epsffile{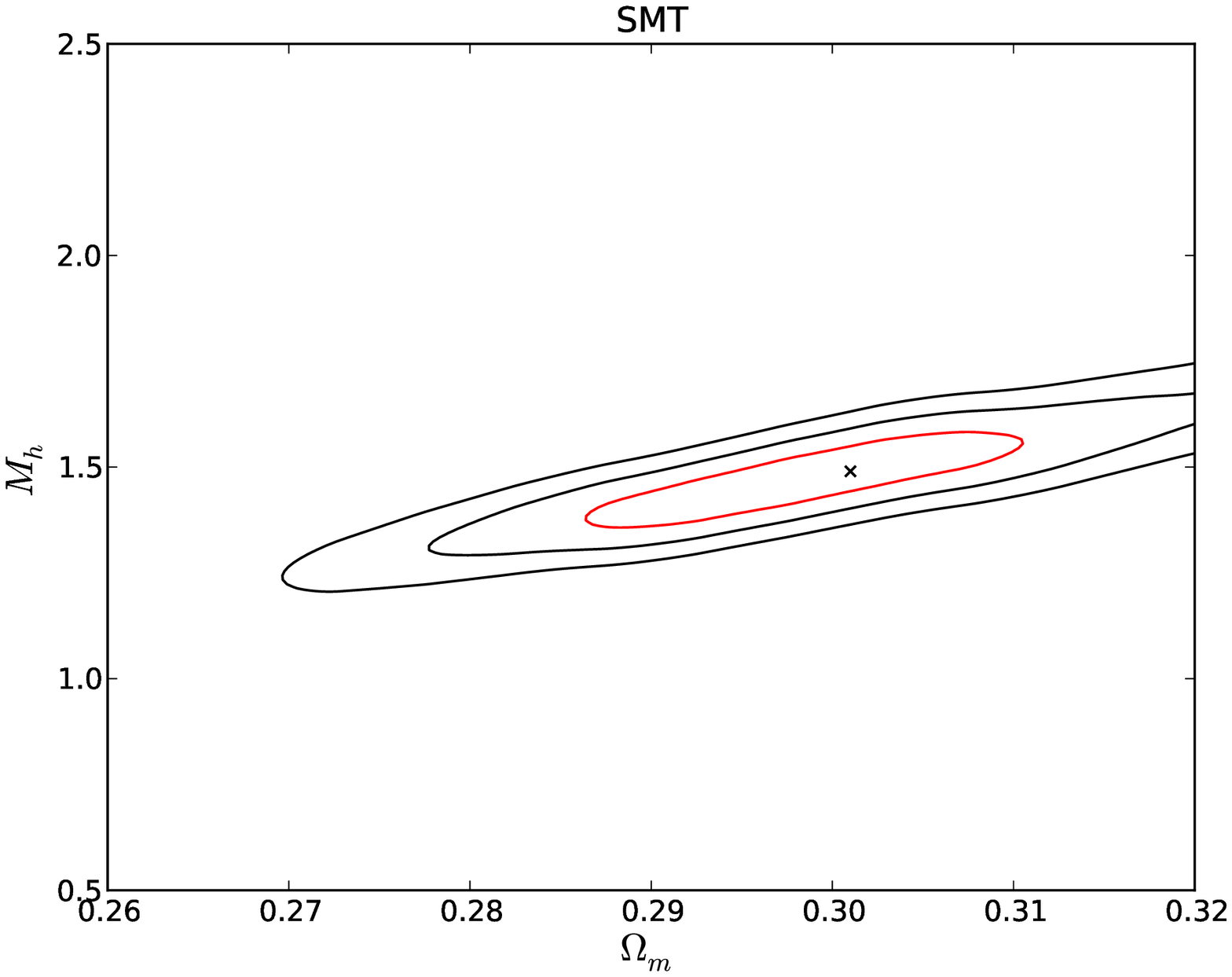}} 
\hfill
\mbox{\epsfxsize=8.2cm \epsffile{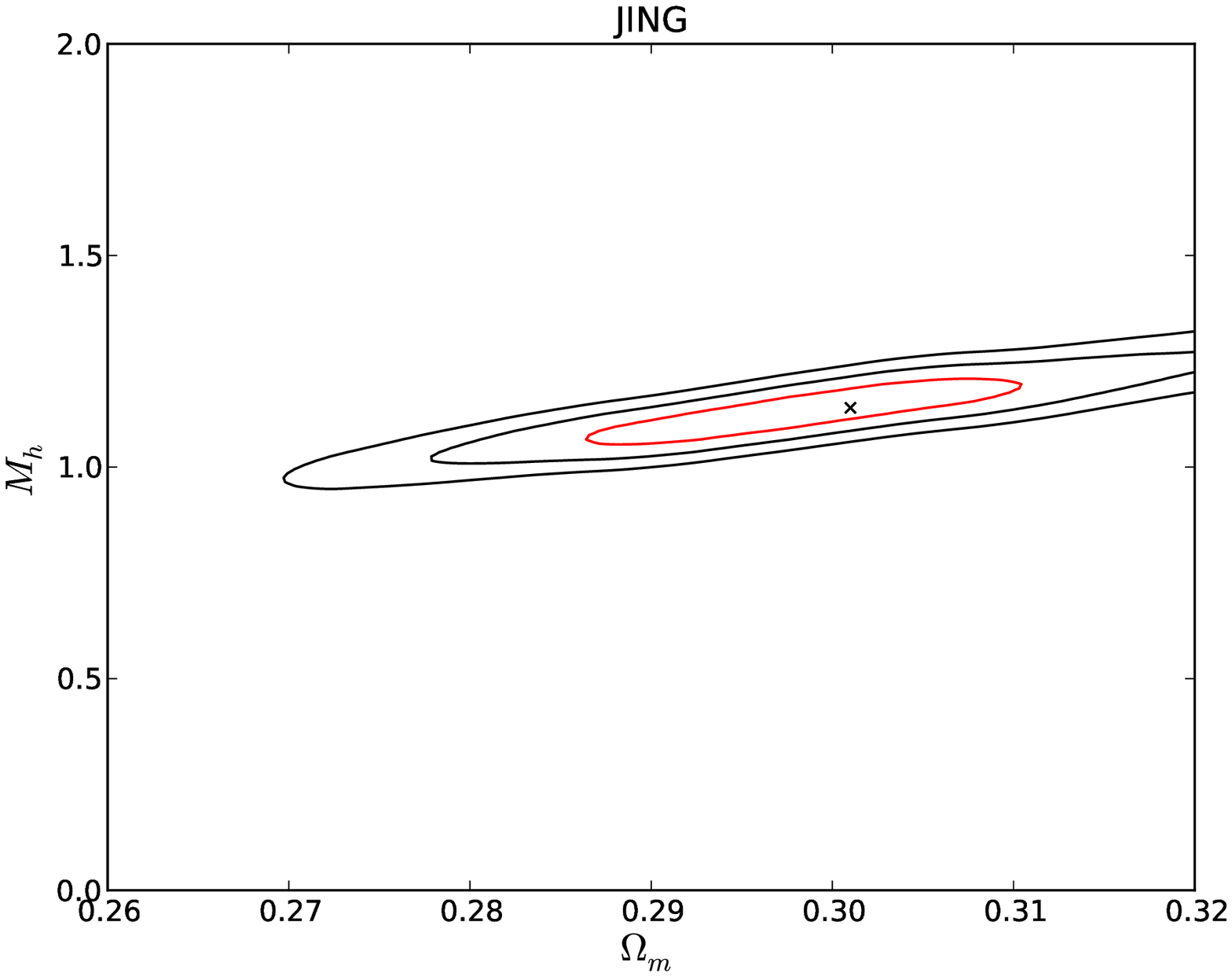}} 
\hfill
\mbox{\epsfxsize=8.2cm \epsffile{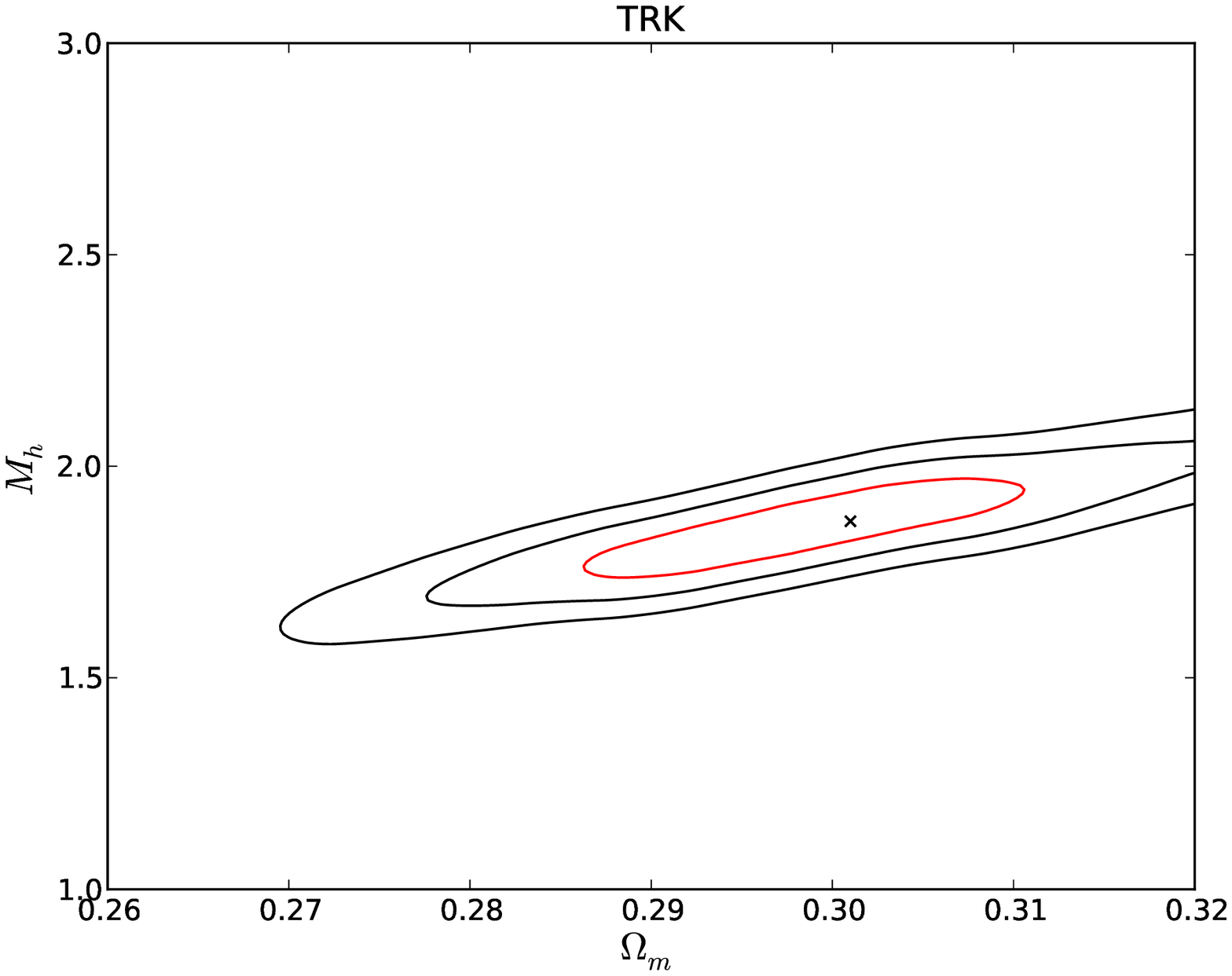}} 
\hfill
\mbox{\epsfxsize=8.2cm \epsffile{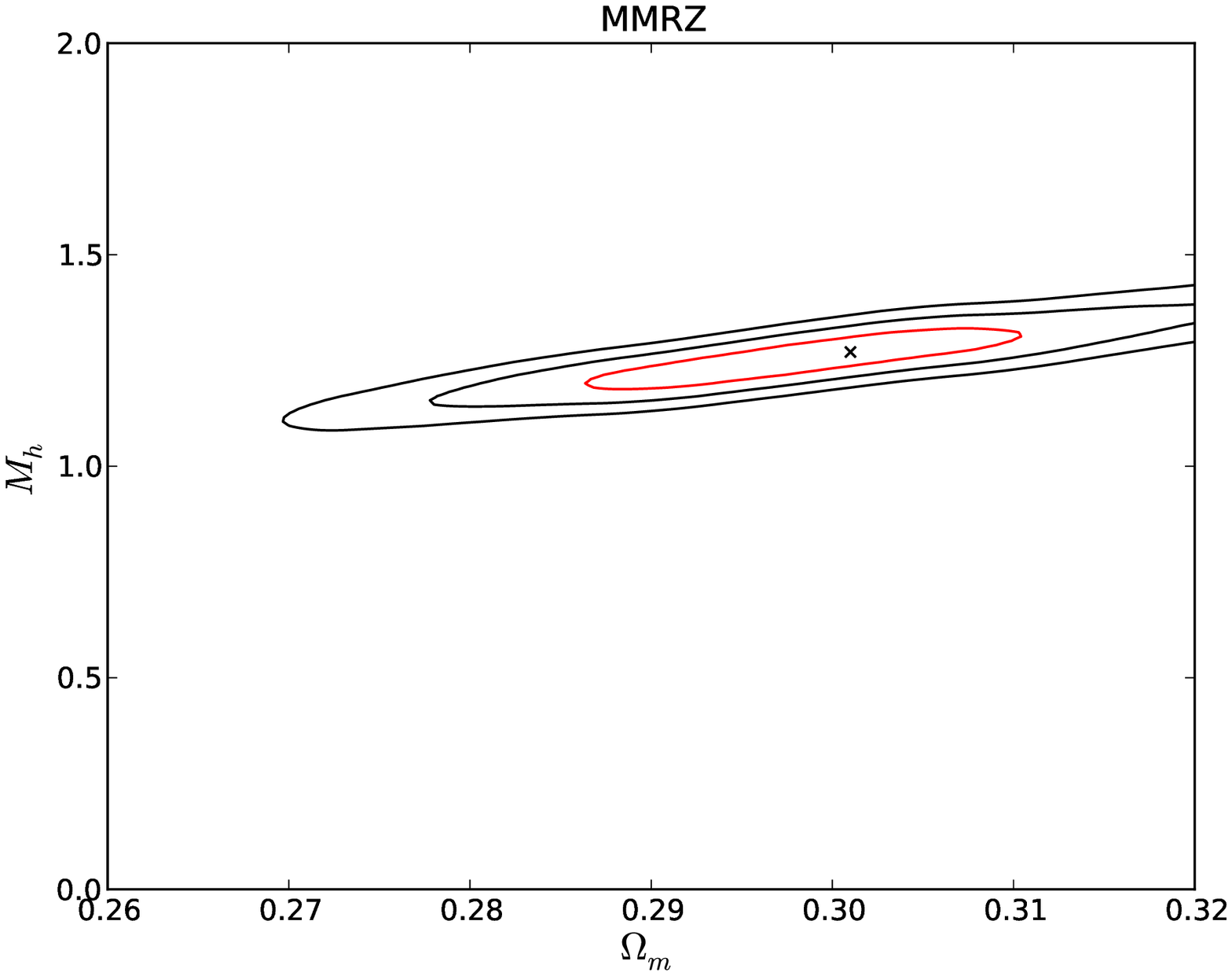}} 
\hfill
\mbox{\epsfxsize=8.2cm \epsffile{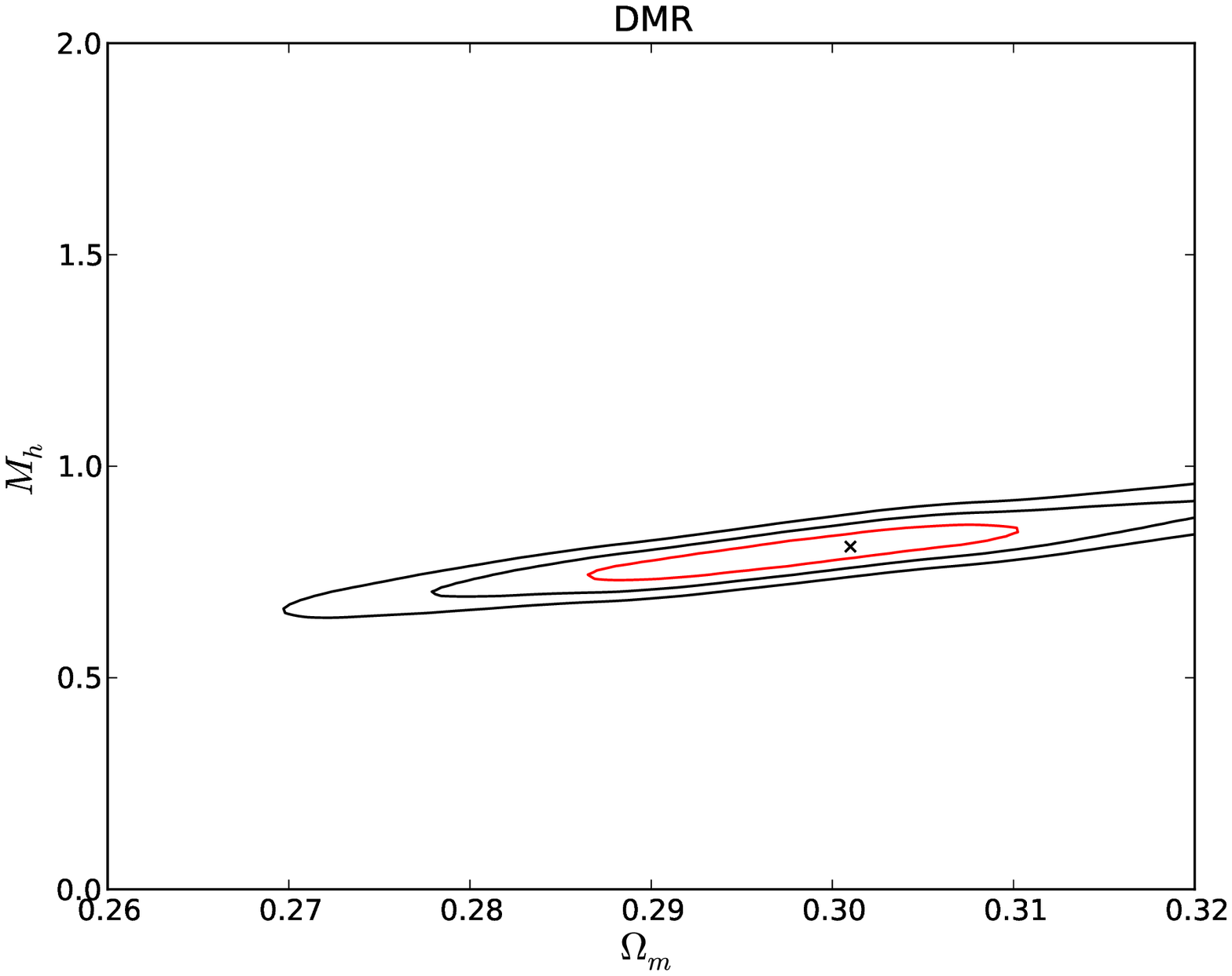}} 
\hfill
\mbox{\epsfxsize=8.2cm \epsffile{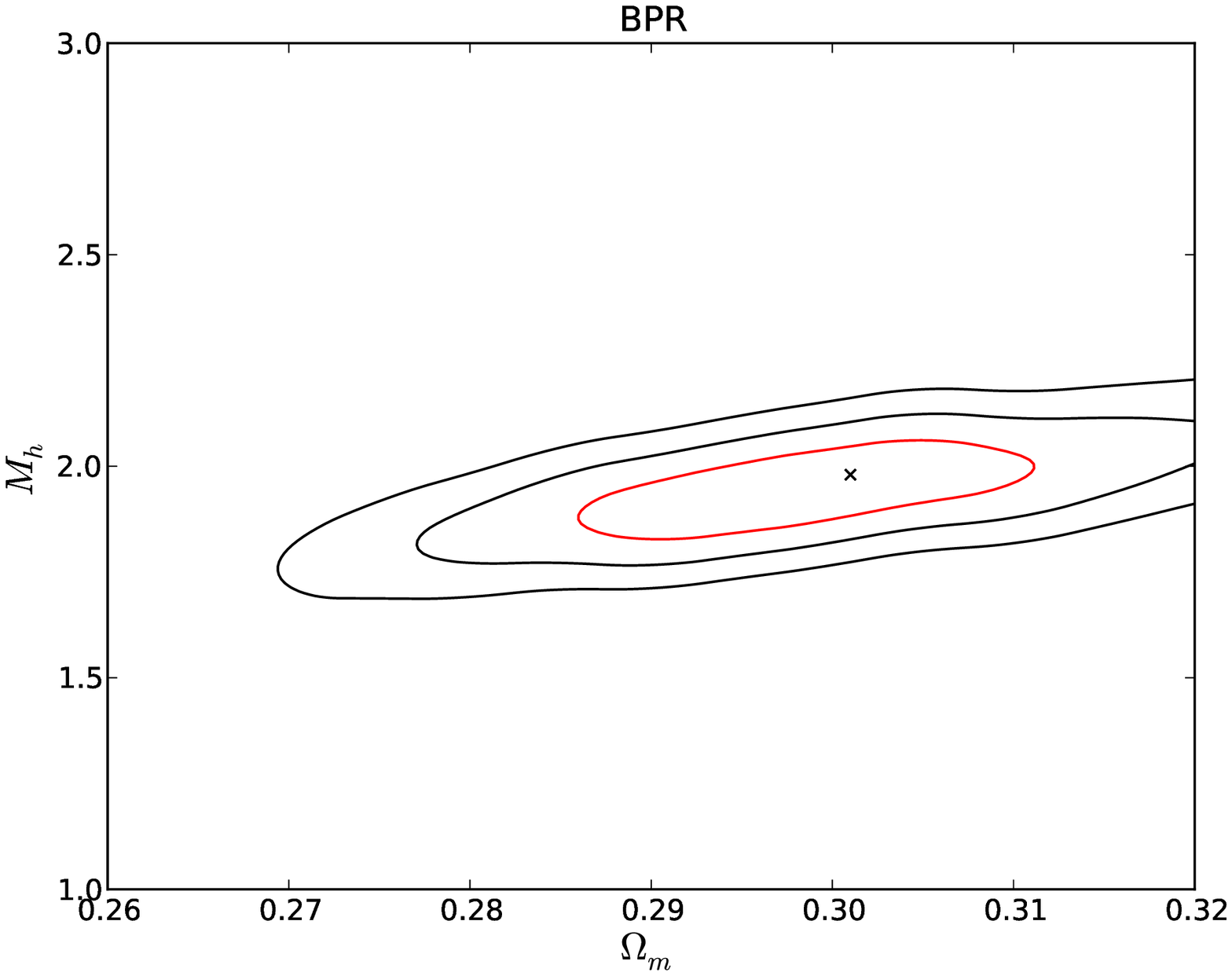}} 
\caption{Contour plot of the two fitted parameters, halo mass, $M_h$
  and $\Omega_m$ for the indicated bias models. 
The $1\sigma$ level is indicated by the red curve.}
\label{f9}
\end{figure*}

In Table \ref{tab6} we present the 
best fit $(\Omega_{m}h, M_{h})$ parameters for the different models and
as it can be seen, the fit to the data, as indicated by the value of
the $\chi^2_{min}$, is equally good for all models, i.e.,
$\chi^{2}_{min} \simeq 7.1$ (with AIC$_{c} \simeq 10.43$) 
for 9 degrees of freedom for all models.
The first interesting result, as can be seen from the Table 3, is
that for all bias models
the likelihood analysis peaks at $\Omega_{m}h=0.204\pm 0.0075$
which reduces to $\Omega_{m}=0.30\pm 0.01$ for $h=0.678$.
Notice, that the latter value
is in excellent agreement with that of 
Planck (TT+lowP+lensing; Ade et al. 2016), of
DES/Planck/JLA/BAO (Abbott et al. 2017) and the re-analysis of the Planck
data provided by Spergel et al. (2015). Using the Planck value of the 
Hubble constant $H_{0}=67.8$km/s/Mpc, provided by Spergel et al. (2015),
in Figure \ref{f9} we present the 1$\sigma$, 2$\sigma$ and $3\sigma$ contours 
in the $(\Omega_{m}, M_{h})$ plane for the SMT, JING, TRK, MMRZ, DMR and BPR
bias models, respectively. 

The second result worth mentioning is that the fitted DM halo mass is
somewhat larger with respect to that of the DES data, namely it ranges between 
$\sim 0.8 \times 10^{13}h^{-1} M_{\odot}$ and $\sim 2 \times
10^{13}h^{-1} M_{\odot}$.
The corresponding weighted mean and the combined weighted
standard deviation of the DM halo mass are:
$${\bar M}_{h}=1.426 (\pm 0.405)\times 10^{13}h^{-1}M_{\odot}$$
As we have already found previously using the DES bias data, also here
the weighted mean DM halo mass tends to that of SMT.

The latter results, concerning the weighted mean DM halo
mass, although based on the integrated clustering of LRGs in one
overall redshift bin, are consistent, within one standard deviation,
to those of Section 4.
This should have been expected since both surveys (DES and 2SLAQ)
trace the same extragalactic objects (LRGs) in a similar redshift range. 
Furthermore, using the TRK bias model we find now that
the derived DM halo mass is consistent at $\sim 2\sigma$ level 
with that of Sawangwit et al. (2011), namely
$M_{h}\simeq 2 (\pm 0.1) \times 10^{13}h^{-1}M_{\odot}$ 
(see also Pouri et al. 2014). 
Notice, that the latter holds for 
the BPR bias model in the case of DES/Planck/JLA/BAO rescaled bias data.
Sawangwit et al. (2011) used the bias model of 
Sheth et al. (2001) together with the revised parameters of 
Tinker et al. (2005). Using these parameters for the SMT 
model in our analysis we obtain
$M_{h}=2.24 (\pm 0.1) \times 10^{13}h^{-1}M_{\odot}$.

Lastly, we verify that perturbations around the values 
$h=0.678$, $n=0.9671$ and $\Omega_{b}=0.0484$ 
do not really affect the aforementioned statistical results.

In Figure 5, we plot the observed $w(\theta)$ for the 2SLAQ
LRGs, with the best fit model of the angular correlation function provided
by Eq.(\ref{eq:wp4}) and the minimization procedure presented above.
Notice, that the solid curve corresponds to the bias model of SMT
with $\Omega_{m}=0.301$ and $M_{h}=1.495\times 10^{13}h^{-1}M_{\odot}$.
The red stars indicate the ACF data in the range $0<\theta <140^{''}$,
which as we have already mentioned in section 5.1 have been excluded 
from our analysis in order to 
avoid the strong non-linear effects. 
We remind the  reader that these scales, 
at the median redshift of $z_{\star}=0.55$, correspond
to spatial separations $\lesssim 1 h^{-1}{\rm Mpc}$.

\begin{table}
\begin{center}
    \begin{tabular} {c  c  c  c  }
    \hline 
    \\
   Bias Model & $10^{13}h^{-1}M_{\odot}$ & $\Omega_{m}h$ &$\chi^2_{min}/d.o.f.$   \\[1.ex]  \hline 
    SMT & $1.495^{+0.075}_{-0.060}$ &$0.204\pm 0.0068$ & $7.05$         \\ [1.ex]
    JING & $1.155^{+0.060}_{-0.040}$ &$0.204\pm 0.0075$& $7.05$         \\ [1.ex]
    TRK & $1.880^{+0.090}_{-0.060}$& $0.204\pm 0.0075$ & $7.05$       \\[1.ex]
    MMRZ & $1.270^{+0.070}_{-0.035}$& $0.204\pm 0.0075$ & $7.06$    \\[1.ex]
    DMR & $0.810^{+0.045}_{-0.030}$ &$0.204\pm 0.0075$& $7.05$     \\ [1.ex]
    BPR & $1.945^{+0.085}_{-0.115}$ &$0.204 \pm 0.0075$& $7.05$     \\ \hline 
    \end{tabular}
\end{center}
\caption {Results of the $\chi^2$ minimization procedure using
the LRGs ACF of 2SLAQ (see Sawangwit et al. 2011 and 
Table 1 in Pouri et al. 2014).
In this case we have $\Delta$AIC $\simeq 0$. 
}
\label{tab6}
\end{table}

\begin{figure}
\mbox{\epsfxsize=8.2cm \epsffile{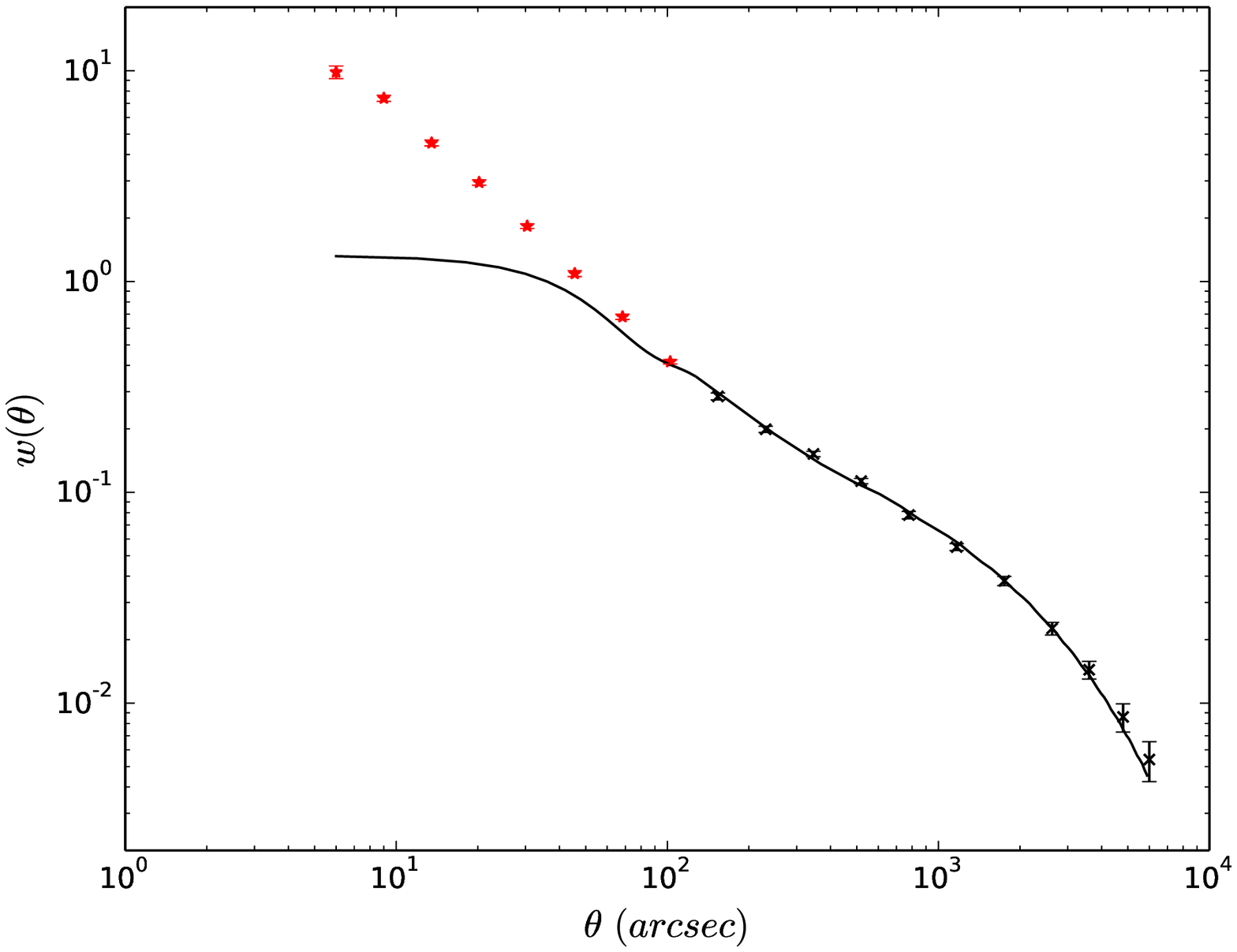}} \caption{Comparison 
between the predicted angular correlation function for the
BPR bias model (solid line) 
and the LRGs correlation function data points of 
Sawangwit et al. (2011) (by red stars we
indicate the non-linear scales: $0<\theta\le 140^{''}$). 
The solid curve corresponds to the
$\Lambda$CDM model expectations using the 
SMT bias model with 
$\Omega_{m}=0.301$ and $M_{h}=1.495\times 10^{13}h^{-1}M_{\odot}$.}
\end{figure}

\section{Conclusions}
In this work we have used the bias data of Luminous Red Galaxies, 
recently released by the Dark Energy Survey (DES) team, in order to 
investigate the ability of six bias evolutions 
models to represent the observational data. Implementing a standard 
$\chi^2$ minimization procedure 
between the models and the data we have placed tight constraints on the
only free parameter of the model, namely the dark matter halo mass
$M_{h}$. Based on the best-$\chi^{2}$ 
and the value of the Akaike information criterion we found that 
most bias models fit equally well the DES bias data. The intrinsic 
differences of the bias models appear to be absorbed in the fitted
value of the dark-matter halo mass in which LRGs live, and which ranges 
between $\sim 6 \times 10^{12}h^{-1} M_{\odot}$ and $1.4 \times
10^{13}h^{-1} M_{\odot}$ for the different bias models. 

The bias model that best fit the DES bias data is that of Sheth et al. (2001), 
while we found indications for a tension between the model of Ma
et. al. (2011) and the bias data.
Moreover, we have shown that the Jing (1998), Tinker et. al. (2010), 
Ma et al. (2011) and the 
Basilakos et al. (2011) bias models predict 
consistent values (within $1\sigma$) of the mass of dark matter halos
hosting LRGs 
with that of Sheth et al. (2001). We have also found that 
regardless the value of the fitted 
DM halo mass, the bias model of Sheth et al. (2001) 
is statistically equivalent to those 
of Jing (1998), Tinker et. al. (2010) and 
de Simone et. al. (2011), since $|\Delta$AIC$|\le 2$.

In the second part of the paper we have used again a standard $\chi^2$
minimization procedure between the theoretical angular clustering
models, which also include the evolution of bias, and the corresponding
2SLAQ LRG clustering.
This analysis also has showed that the bias models explored 
are statistically equivalent.
Furthermore, it provided a value of $\Omega_{m}=0.30\pm 0.01$,
which is in excellent agreement with that of 
Planck (TT+lowP+lensing; Ade et al. (2016), 
DES/Planck/JLA/BAO (Abbott et al. 2017) and the reanalysis of the Planck
data (Spergel et al. 2015). 
Finally, concerning the estimated DM halo mass, the clustering
analysis has provided a range between $8 \times 10^{12} h^{-1}
{\rm Mpc} \lesssim M_{h} \lesssim 2 \times 10^{13} h^{-1} M_{\odot}$,
results which are somewhat larger with those based on the DES
  bias data. However, 
using an inverse-AIC$_c$ weighting we find that the model average value of the
DM halo mass that hosts LRGs are consistent within 1$\sigma$ using
either the DES or the SDSS 2SLAQ analyses.

\section*{Acknowledgments}
S. Basilakos acknowledges support by the Research Center 
for Astronomy of the Academy of Athens in the
context of the program ''Testing general relativity on cosmological scales''
(ref. number 200/872).

\end{document}